# Three-dimensional broadband FDTD optical simulations of CMOS image sensor

A. Crocherie[a,b], J. Vaillant[a], and F. Hirigoyen[a]
[a] STMicroelectronics, 850 rue Jean Monnet, 38920 Crolles, France ;
[b] IMEP-LAHC, MINATEC, 3 parvis Louis Néel, 38016 Grenoble, France

## ABSTRACT

In this paper, we present the results of rigorous electromagnetic broadband simulations applied to CMOS image sensors as well as experimental measurements. We firstly compare the results of 1D, 2D, and 3D broadband simulations in the visible range (380nm-720nm) of a 1.75µm CMOS image sensor, emphasizing the limitations of 1D and 2D simulations and the need of 3D modeling, particularly to rigorously simulate parameters like Quantum Efficiency. Then we illustrate broadband simulations by two proposed solutions that improve the spectral response of the sensor: an antireflective coating, and the reduction of the optical stack. We finally show that results from experimental measurements are in agreement with the simulated results.

## 1. INTRODUCTION

CMOS Image sensors market has experienced an important growth over recent years due to the increasing demands of video cameras, security cameras, webcam, and mainly mobile cameras [1,2]. This evolution leads to a race to miniaturization by the main market's protagonists: pixel size shrinks because the resolution (number of pixels) increases while keeping small sensors. Nevertheless, there is a drawback to this increase of resolution: as photodiode area shrinks, less photons are collected and the sensibility of the detector is reduced. Besides, as pixel size decreases, diffraction effects caused by metallic interconnection are emphasized, and the amount of signal collected by adjacent photodiodes (known as cross-talk) increases.

In order to optimize sensors performances and anticipate potential optical limitations, rigorous optical simulations are essential. For pixels smaller than 2µm, ray-tracing models [3] become inaccurate and a more fundamental description based on Maxwell-Boltzmann modeling is an alternative to simulate optical propagation in CMOS image sensors [4]. We chose an electromagnetic simulation tool based on Finite Difference Time Domain (FDTD) method [5,6], available from Lumerical Solutions [7], to describe light propagation and photon collection inside the pixels while correctly simulating diffraction effects. We also developed with this software a method to simulate a diffuse-like source that reproduces real illumination conditions [8]. We will see in this paper the importance of 3D broadband simulations to optimize these sensors, and will show some applications in order to improve performances of CMOS Image sensors.

The paper is organized as follows: in section 2 we will present briefly the FDTD method and describe the general concepts to realize broadband simulations with Lumerical software. In section 3 geometrical effects on 1D, 2D, and 3D broadband simulations will be presented. Section 4 deals with two applications for optimization of the sensor performances, an antireflective coating and the stack reduction. Finally, section 5 will compare results from simulation to Quantum Efficiency measurements.

## 2. INTRODUCTION TO BROADBAND SIMULATIONS

### 2.1 Finite Difference Time Domain method

The FDTD method has become the state of the art to solve electromagnetic problems with complex geometries. It has been applied many times to simulate structures in the fields of photonics and optics and more specifically to simulate pixels in the CCD [9] and CMOS [10-13] technologies.

The implementation is discrete in both space and time. The time step is selected to ensure numerical stability and is related to the mesh size through the speed of light [6]. Structures of interest are described on a discrete mesh made up of Yee-cells [5] and Maxwell's equations are solved discretely in time on this mesh:

$$\frac{\partial \vec{E}}{\partial t} = \frac{1}{\varepsilon}\left(\vec{\nabla} \times \vec{H} - \sigma \vec{E}\right) \quad (1)$$

$$\frac{\partial \vec{H}}{\partial t} = -\frac{1}{\mu_0}\left(\vec{\nabla} \times \vec{E}\right) \quad (2)$$

with $E$, electric field, $H$ magnetic field, $\varepsilon$ permittivity (see section 2.3), $\mu_0$ permeability of the vacuum (non magnetic material here) and $\sigma$ conductivity of the medium.

We can see that any change in the E-field in time is linked to any change in the H-field across space and vice versa, which is the basis of the FDTD implementation. Finally, FDTD is a fully vectorial method that gives both time-domain and frequency domain information by exploiting Fourier Transforms; i.e. when a broadband pulse is used as the source, the response of the system over a wide range of wavelengths could be obtained in a single simulation, so that many quantities could be obtained like the complex Poynting vector or the transmission/reflection of the light.

## 2.2 Accuracy of simulation

The mesh step affects the accuracy of the simulation: a smaller mesh step will give more accurate results but requires more resources. One defines a parameter $N_\lambda$ that represents the number of mesh cells per wavelength as shown by equation (3):

$$N_\lambda = \frac{\lambda}{\Delta x . n} \quad (3)$$

with $\Delta x$ the mesh step along one direction of the grid, $\lambda$ the smallest wavelength of the simulation source, and $n$ the highest refractive index of the simulation structure.

As $N_\lambda$ increases for $\lambda$ and $n$ given, the mesh step $\Delta x$ decreases and the simulation results are more accurate. As a rule of thumb, for simple structures, the mesh step should be set less than one-tenth of the wavelength in the highest index material. Nevertheless, for structures with complex geometries and wavelength scale feature size, the mesh size must be reduced to capture the index variations. In this aim, we have run convergence test to determine what setting is required to ensure accurate results.

This test consists in simulating the transmission and reflection of light, respectively at photodiode interface and above the optical stack, of one 2D CMOS sensor pixel for one plane wave at normal incidence from 380nm to 720nm. Perfectly Matched Layer (PML) absorbing boundary conditions are used on top and bottom of the structures, and Bloch periodic boundary conditions are used on the sides of the structures. More explanations on pixel structure and transmission calculations will be detailed in section 2.4. In this test, all parameters are fixed except the mesh step which changes from 60nm to 1nm corresponding to $N_\lambda$ from 1 to 50 at 380nm in the highest index material (silicon). One has to notice that uniform grid is used in this paper, i.e. the mesh step is imposed in the whole simulation area by the highest index material. Finally, as analytic solutions for the transmission and reflection can't be calculated because of microlens and metal lines, reference for calculation has been taken for the most precise mesh with $N_\lambda=50$. Figure 1 on the left presents the results of transmission and reflection from 380nm to 720nm for four different $N_\lambda$, and right part presents the maximum error on transmission and reflection (on the whole broadband spectrum) compared to the reference $N_\lambda=50$ for different $N_\lambda$.

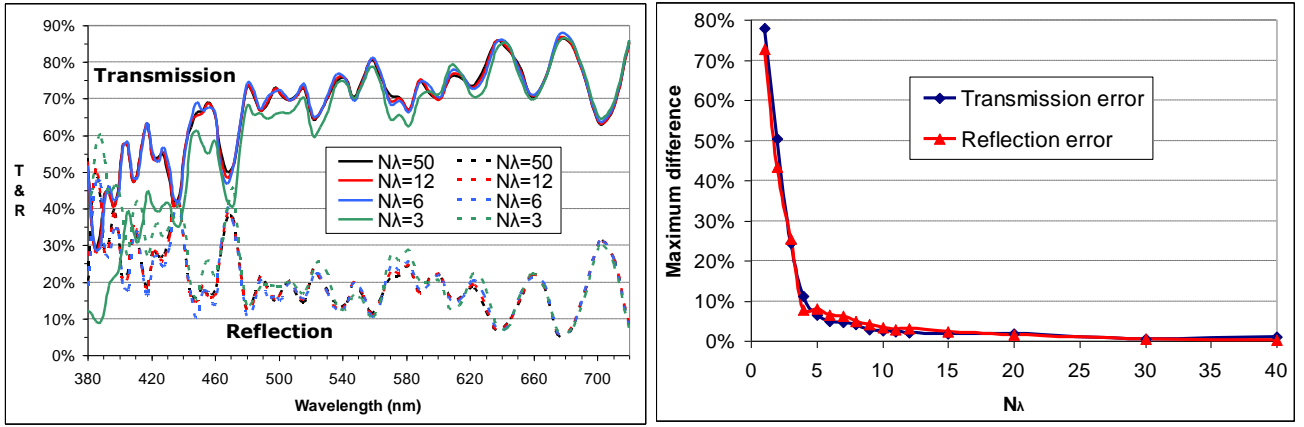

Fig. 1. Transmission and reflection versus wavelength (left) and maximum error versus $N_\lambda$ (right)

On the left plot of Figure 1, we see that when $N_\lambda$ decreases, the transmission and reflection results become less accurate compared to the reference result with the finer grid ($N_\lambda$=50) and this difference is more important for the shortest wavelengths. Studying the plot on the right, we see that convergence is reached for about $N_\lambda$>10, i.e. a mesh step of 6nm at 380nm in silicon. For this accuracy, error is less than 3%. We have also checked that results are not different for 3D simulations. This error could be decreased either by increasing the mesh accuracy, or by increasing the number of PML layers to improve the absorbing boundary conditions, nevertheless both solutions require more resources mainly for complex 3D structures.

In the simulations presented in this paper, the sampling grid has been fixed to 10nm, which corresponds to $N_\lambda$=6 in the silicon at 380nm. This value is a good trade-off between mesh accuracy and resources requirement for 3D simulations. For example, the project simulation size for a 1.75µm pixel with $N_\lambda$=10 (respectively $N_\lambda$=6) is about 33 MB (16 MB) in 2D and 12 GB (3 GB) in 3D leading to a time of simulation of 14 minutes (3 minutes) in 2D and 106 hours (17 hours) in 3D on a Linux workstation with 40 GB of RAM using 1 processor (32 bits Pentium IV 3 GHz). For this value the error is a bit more important, about 5% for the smallest wavelength, but as shown in Table 1 for the wavelengths of interest in CMOS image sensor (from 450nm to 630nm), relative accuracy is better and closed to the one estimated for the convergence of the results. Besides, region that is critical to obtain a good transmission at silicon interface corresponds to the back-end where oxides have lower refractive index. Thus, with a uniform grid fixed at 10nm, $N_\lambda$=6 in silicon at 380nm but is equivalent to $N_\lambda$=26 in oxide, a very precise mesh sufficient to ensure accurate results (see Table 1).

Table. 1. Refractive index and relative mesh accuracy for different material and wavelengths for a mesh step of 10nm

| $\lambda$ (nm) | | 380 | 450 | 530 | 630 | 720 |
|---|---|---|---|---|---|---|
| silicon | n | 6.5 | 4.7 | 4.2 | 3.9 | 3.8 |
| | $N_\lambda$ | 5.8 | 8.0 | 12.6 | 16.2 | 18.9 |
| oxide | n | 1.45 | 1.45 | 1.44 | 1.44 | 1.44 |
| | $N_\lambda$ | 26.2 | 31.0 | 36.8 | 43.8 | 50 |
| air | n | 1 | 1 | 1 | 1 | 1 |
| | $N_\lambda$ | 38 | 38 | 38 | 38 | 38 |

### 2.3 Material properties

To characterize CMOS image sensors, broadband simulations are realized in the visible range from 380nm to 720nm. Unlike monochromatic simulations that need only the material property (refractive index n+ik) at the wavelength of interest, broadband simulations require that refractive indexes are fitted on a dispersive model. Dispersive models available in Lumerical are Debye, Lorentz and Plasma defined in equations (4), (5), and (6) respectively [6,7]. They result in a complex-valued permittivity which varies as a function of the frequency $f$:

$$\tilde{\varepsilon}_D(f) = \frac{\varepsilon_{DEBYE} \cdot \upsilon_c}{\upsilon_c - i2\pi f} \quad (4)$$

$$\tilde{\varepsilon}_L(f) = \frac{\varepsilon_{LORENTZ} \cdot \varpi_0^2}{\varpi_0^2 - 2i\delta_0 2\pi f - (2\pi f)^2} \quad (5)$$

$$\tilde{\varepsilon}_P(f) = \frac{\varpi_p^2}{2\pi f \cdot (i\upsilon_c + 2\pi f)} \quad (6)$$

where $\varepsilon_{DEBYE}$ (Debye permittivity), $\upsilon_c$ (collision frequency), $\varepsilon_{LORENTZ}$ (Lorentz permittivity), $\omega_0$ (undamped resonant frequency of the medium), $\delta_0$ (damping coefficient), and $\omega_p$ (plasma frequency) are the different coefficients to be determined.

The total, complex-valued permittivity of the materials is given in equation (7) where each component arising from the different dispersive models plays a role in the overall permittivity:

$$\tilde{\varepsilon}(f) = \varepsilon_{real} + i \cdot \varepsilon_{imag} \cdot \frac{f_{sim}}{f} + i \cdot \frac{\sigma}{2\pi f \varepsilon_0} + \tilde{\varepsilon}_D(f) + \tilde{\varepsilon}_L(f) + \tilde{\varepsilon}_P(f) \quad (7)$$

The first two terms represent the contribution due to the background permittivity $\varepsilon$, the third term represents the contribution due to the conductivity $\sigma$, and the fourth through sixth terms represent the Debye, Lorentz and Plasma contributions discussed previously. Besides, $f_{sim}$ is the center frequency of the simulation [7].

With this model, the response of the system over the whole visible range can be obtained with a single simulation thanks to the time domain technique FDTD.

### 2.4 Transmission calculation

A CMOS sensor consists of an array of pixels on a silicon wafer, each containing a photodiode where photons are collected. Then several transistors made the collection and the reading of the photogenerated electrons. Above the photodiode, we can find metal layers for interconnections separated by dielectric layers, passivation layers, color filters for color reconstruction, and microlenses on top of the stack to focus light on each photodiode (see Figure 2).

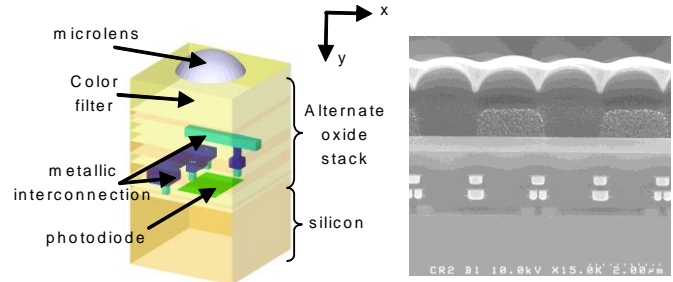

Fig. 2. CMOS image sensor: schematic (left) and SEM picture (right) of 1.75μm CMOS pixel from STMicroelectronics

There exist optical losses in the stack like losses by reflection and diffraction due to interconnect metal layers and losses by absorption due to some dielectric layers. Then, we can firstly define Optical Quantum Efficiency (OQE) which is the ratio of the photons hitting the photodiode area over the incident photons. Thus OQE could be calculated by the Transmission throughout the stack until silicon interface as in equation (8) for 2D simulation for a plane wave propagating in y-direction:

$$OQE(f) = T_y(f) = \frac{\int P_y(\text{silicon interface}).dx}{\int P_y(\text{source}).dx} \qquad (8)$$

with the Poynting vector $P$ in W.m$^{-2}$.Hz$^{-2}$ calculated by the FDTD simulation and defined in equation (9):

$$\vec{P}(f) = \frac{1}{2}.\text{Re}\left[\vec{E}(f) \times \vec{H}(f)\right] \qquad (9)$$

with $E$ and $H$ the electric and magnetic field respectively in V.m$^{-1}$.Hz$^{-1}$ and A.m$^{-1}$.Hz$^{-1}$.

Transmission results will be presented in the next section for 1D, 2D, and 3D simulations.

## 3. GEOMETRICAL EFFECTS ON BROADBAND SIMULATIONS OF CMOS SENSOR

In this section, we present the effects of 1.75µm pixel architectures on broadband simulations in the visible range. We simulate the pixel from the simplest structure in One-Dimension to the Two-Dimensional modelization and finally the real Three-Dimensional pixel.

### 3.1 1D Broadband simulations

The first work is the simulation of the sensor stack in One-Dimension, i.e. only the stack of dielectric and passivation layers is present without metallic interconnections and microlenses. This first approach is the standard one in CMOS sensor simulations when different optical stacks have to be compared in term of transmission. Simulation is made with the FDTD software (electromagnetic calculation) and transmission at silicon interface is compared to the transmission determined by classical 1D thin-films matrix formalism [14]. Both results simulated at normal incidence from 380nm to 720nm, with the materials parameterized according to the dispersive models seen on section 2.3, are plotted on Figure 3.

Transmission plots show good agreement between both simulations. Oscillations are caused by the interferences in the sensor stack. The difference between both simulations, in addition to the calculation method of course, is that in the matrix formalism we consider the silicon substrate as semi-infinite whereas in electromagnetic simulations its dimension is finite fixed by the grid simulation area and bounded by absorbing conditions at the bottom here. Besides, in FDTD, layers dimensions are defined by the mesh step and thus are not exactly the same as in the 1D matrix formalism.

We have also plotted on Figure 3 the transmission calculated by the thin-films matrix formalism with the experimental materials: the result is very close to the other plots showing the good parameterization of the different materials with the models defined in section 2.3.

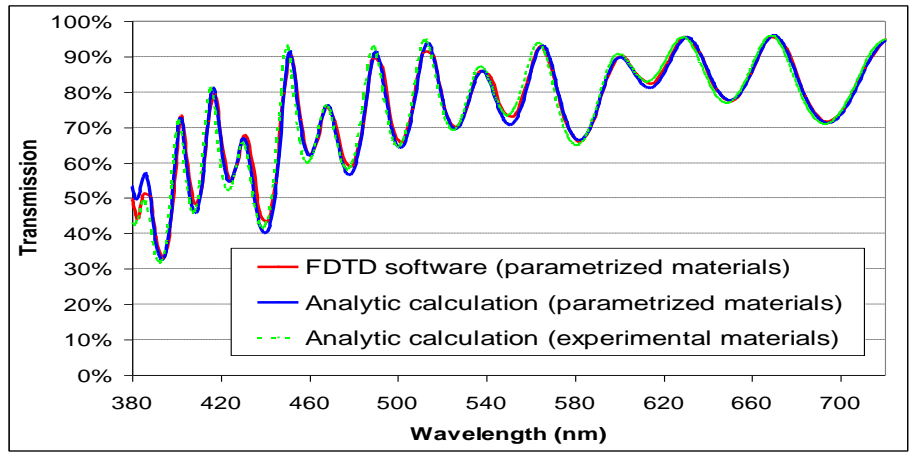

Fig. 3. Transmission simulated at normal incidence at silicon interface as a function of wavelength

## 3.2 2D Broadband simulations

The second step was to add the different elements in the stack that cause the optical losses, i.e. the metallic interconnection responsible for diffraction and reflection of the light, but also the polysilicon gate of the transistor at silicon interface responsible of absorption of the light.

Three different stacks are compared here and detailed in Figure 4. The first one is the previous stack with only uniform layers. In the second one diffractive elements of the pixels have been added, and in the third one the final pixel is simulated by adding the microlenses on top of pixels. One has to notice that color filters have been removed on the pixels and replaced by equivalent planar layer thickness, to allow a simulation on the whole visible range keeping constant the optical stack height of the sensor. Pixels simulated here have a pitch of 1.75µm and the height of the stack is around 3.3µm.

These three pixels are simulated with FDTD method and transmission results at silicon interface are plotted on Figure 5. The source used is a plane wave at normal incidence from 380nm to 720nm with TE/TM polarization average. The transmission is calculated here on the whole pixel surface.

As we could expect, the stack with the uniform layers plus the diffractive elements presents the worst transmission due to the losses caused by the metallic interconnection and the polysilicon gate. With the microlenses the transmission increases thanks to the focalization of the light and losses are reduced. Besides, the microlenses allow a smoothing of the interferences in the stack as we could see on Figure 5.

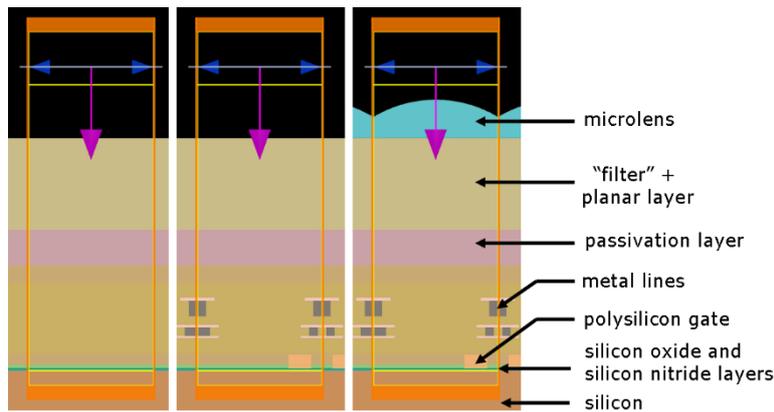

Fig. 4. Layout of the different pixels simulated. Uniform layers (left), uniform layers and diffractive elements (center), final pixel with microlens (right). Color filters have been replaced by equivalent planar layer thickness.

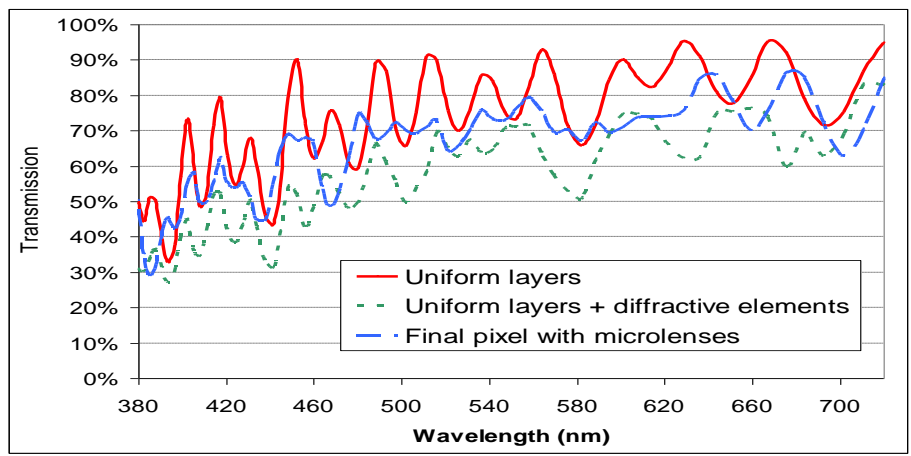

Fig. 5. 2D Transmission simulated at normal incidence at silicon interface as a function of wavelength for the three different configurations described in Figure 4.

## 3.3 3D Broadband simulations

Finally, the same study is made here between the three different stacks but for 3D simulations at normal incidence. A 3D view is represented on Figure 6 with diffractive elements visible (microlenses, metal lines, polysilicon gate) and results are plotted on Figure 7.

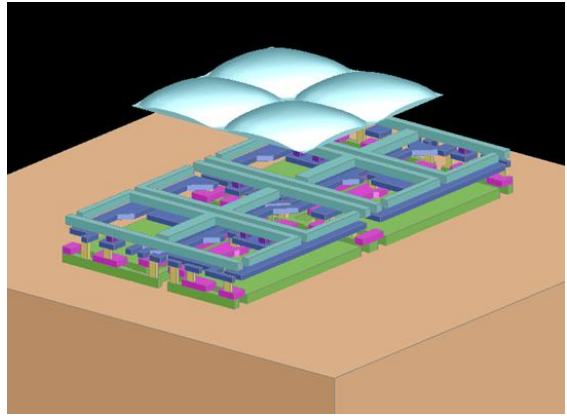

Fig. 6. 3D view of the real monochrome stack simulated

We could first check that the transmission of the uniform layer pixel is the same as the one simulated in 1D and 2D. The other plots present not only a diminution of their average transmission compared to the 2D simulation, but also a variation in the oscillations. These observations could be explained simply: in 2D simulation metal lines are considered as infinite in the third dimension whereas in the real case, i.e. in 3D metal lines define an aperture of finite dimensions centered on the photodiode of each pixel as shown on Figure 6. Thus, losses are under-evaluated in 2D simulations. Moreover, as pixel size decreases, in order to keep a photodiode large enough to collect photons, some transistors are shared between several pixels, which leads to asymmetric architecture of pixels [1]. This asymmetry could not be simulated in 2D, which is the reason why 3D transmission plots present different and more realistic performances than in 2D.

Besides, oscillations due to interferences in the stack clearly show that monochromatic simulations are not adapted to characterize the performances of a CMOS image sensor. Thus, 3D broadband simulations are essential to rigorously determine the characteristics of CMOS sensor like the OQE presented here in the plots.

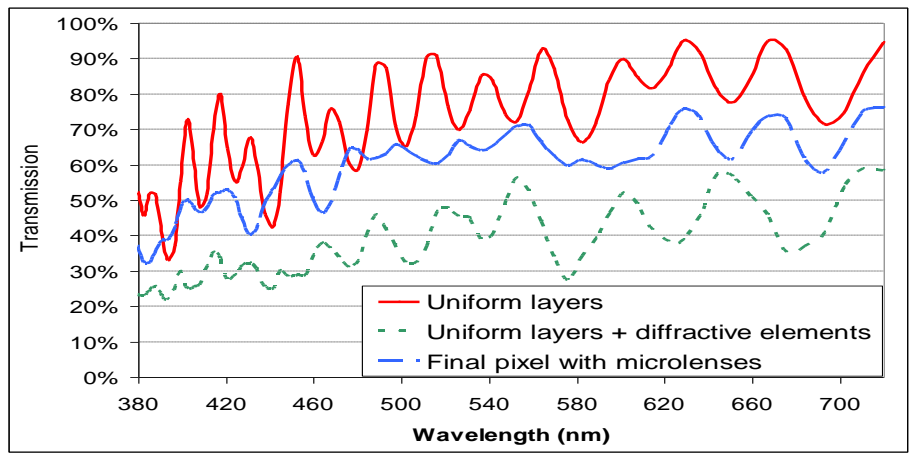

Fig. 7. 3D Transmission simulated at normal incidence at silicon interface as a function of wavelength for the three different configurations described in Figure 4.

We could also see on the plot another important parameter which is the microlens gain (transmission ratio between the stack with microlenses and the stack without). This gain is greater than in two-dimension and varies from 1.5 to 2 depending on the wavelength, which means that most of the losses on the diffractive elements have been cancelled thanks to the focalization of the light.

Finally, one parameter that we have not mentioned yet here is the Fill Factor. In fact, the collection area of photons at silicon interface (i.e. the photodiode) does not fill the whole pixel area; a part is occupied by the transistors for the purpose of collecting, converting and reading photogenerated electrons. The ratio of the photodiode area to the whole pixel area is called the Fill Factor. We have thus plot the same 3D transmission on Figure 8 for a Fill Factor of 50%, a realistic value for that pixel pitch. Compared to the Figure 7, we see that there is less signal for the two first configurations: the uniform layers case and the diffractive case, whereas the final pixel with microlenses presents almost the same performances which means that our microlens is well optimized to focalize the light on the photodiode.

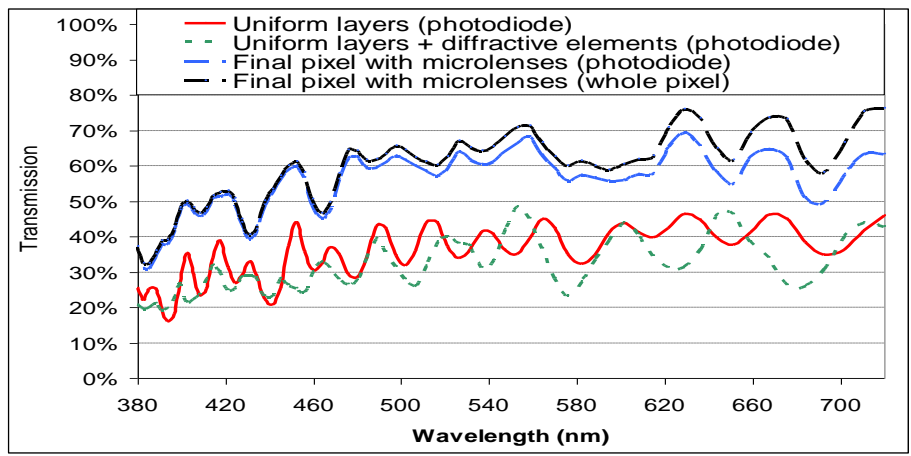

Fig. 8. 3D Transmission simulated at normal incidence at silicon interface as a function of wavelength for the three different configurations described in Figure 4 and with a signal integrated on the photodiode area.

## 4. OPTIMIZATION OF THE SENSOR PERFORMANCES

We will now present in this section two applications of 3D broadband simulations in order to optimize the sensor performances. The sensor studied here is the same as previously, i.e. a pixel with 1.75μm pitch and 3.3μm height from STMicroelectronics [15]. In this section, transmission is calculated for the photodiode area (50 % Fill Factor).

**4.1 Antireflective coating**

The first application is to minimize the reflections at silicon interface. In fact in standard CMOS process, different thin layers (few tens of nanometers) are deposited above the silicon like silicon oxide or silicon nitride for manufacturing process reasons. Nevertheless, thicknesses of these layers are not optimized to sensor problematic where minimization of losses is basic: reflection at silicon interface could be decreased if the thicknesses of these layers are modified to create a best fit between the high refractive index of the silicon and the index of the optical stack layers. In a first step, 1D thin-films matrix formalism allows that kind of optimization: we choose to minimize reflections in the green range around 530nm and this simulation leads to change the silicon nitride layer thickness, for a gain in transmission of about 1.08. Then, with 3D FDTD broadband simulations, we plot in Figure 9 the transmission response of both configurations at normal incidence to verify the gain of the new structure in the green range and to check that it does not decrease the transmission at others wavelengths. The gain of 1.08 agrees with the FDTD simulations in the green range because averaged transmission at 532±10nm changes from about 62% to 67% at normal incidence. Besides, there is no significant change in the red range and transmission in the blue range slightly increases too.

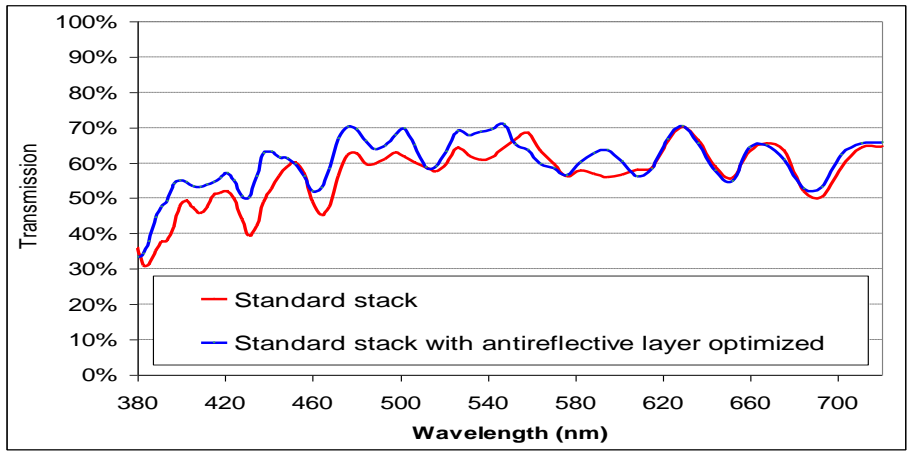

Fig. 9. Transmission simulated at normal incidence at silicon interface as a function of wavelength for two configurations

### 4.2 Stack reduction

The second one consists in a reduction of the optical stack of our sensor. We have seen previously that the optical transmission of the stack is a key parameter in the collection of the photons in the photodiode. While decreasing the stack height and thus the optical path length of the photons, we expect to reduce the losses in the stack and then to increase the transmission at silicon interface (i.e. the sensitivity of the sensor). We proposed to reduce the stack height above the photodiode of approximately 30% by removing layers to get a better flux concentration. We could expect an important gain with a product-like aperture where important incidence angles lead to cross-talk. To evaluate the optical stack height reduction with this product-like aperture, we have performed 3D narrowband simulations at 532±10nm with a diffuse-like source at f#=2.8 (i.e. incidence angle between -10° and +10°) on a real pixel with color filters. Complete methodology used for this kind of simulation has already been published [4,8]. We integrate the signal in a volume defined by the photodiode area (50% Fill Factor) and a depth of 3µm in the silicon. Our electromagnetic simulations evaluated the potential optical gain of this new reduced process at 1.15 for this aperture. Poynting results of the two configurations, the standard one and the reduced stack, is represented in Figure 10 and show a better flux concentration with the reduced stack.

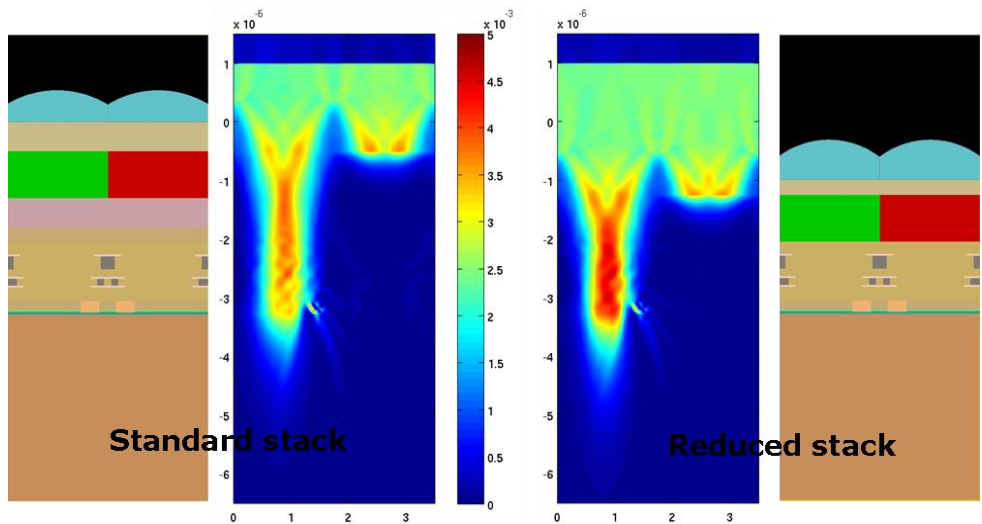

Fig. 10. Poynting results from FDTD simulation with f#=2.8 for the standard and the reduced stack

# 5. COMPARISON WITH MEASUREMENTS ON THE REAL PIXEL

## 5.1 Description of the optical bench for Quantum Efficiency (QE) measurements

The characterization bench for QE measurements is represented on Figure 11. It is composed of a halogen light source and a monochromator in order to scan the visible spectrum from 380nm to 720nm. Then, a beam splitter redirects a small fraction of the light to a calibrated photodiode for real-time light level control. The final part of the bench controls the shape of the illumination: diffuse light with given f-number. This is done by an opal diffuser and the control of the diffuser to sensor distance. If the diffuser diameter is large enough, the size of the sensor can be neglected then we can consider that all pixels inside the sensor see the same illumination shape, defined by the f-number.

Then, standard image post-processing is applied for data at each wavelength (time averaging for temporal noise reduction and dark frame subtraction).

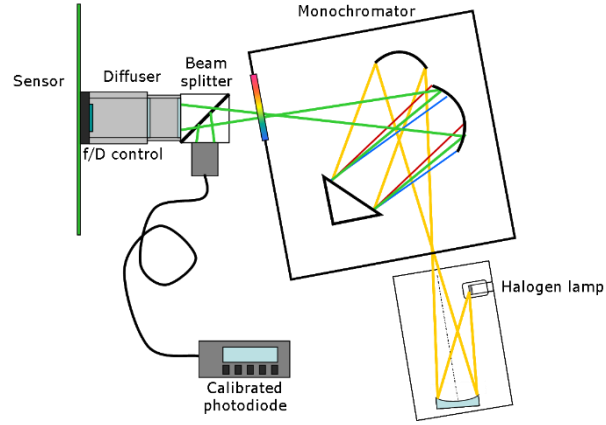

Fig. 11. Typical test bench configuration: sensor to test is uniformly illuminated (placed behind a diffuser).

## 5.2 Validation of the monochrome broadband modelization

One can notice that all the broadband simulations presented here have been made at normal incidence on different stack where color filters have been replaced by equivalent planar layer thickness (exception made on the simulations on Figure 9). This model allows broadband simulations on sensor where the optical stack height and the microlenses are unchanged compared to the final product. In the next section, we will present results from measurements on stack with color filters to see the impact of our optimization on real sensor. Nevertheless, to qualify the relevance of our model studied here on broadband spectrum, we have processed wafers at STMicroelectronics with color filters replaced by equivalent planar layer thickness in order to be exactly in the same conditions as simulation. Both results from simulation and measurements are plotted on Figure 12 and show the good agreement between them. The pixel studied here is the reduced stack configuration at normal incidence.

Besides, as transmission on simulation is calculated at silicon interface, to compare with the measurements, we have to take into account the absorption in the silicon dependent on the wavelengths but also on the different doping profile in the silicon [9]. In our chip, we use a P+/N/P- photodiode on a P+ substrate. We know that some carriers are recombined in the P+ regions but the P+/N and P-/P+ boundaries are not well defined. In our simplified model, we choose to count only the electron-hole pairs generated in the N/P- region, i.e. between two depths in the silicon, $z1$ and $z2$, depending on the process. This leads to equation (8) for the final 3D OQE:

$$OQE(\lambda) = T(\lambda) \cdot \frac{\int_{z1}^{z2} e^{-\alpha \cdot z} dz}{\int_{0}^{\infty} e^{-\alpha \cdot z} dz} = T(\lambda) \cdot (e^{-\alpha \cdot z1} - e^{-\alpha \cdot z2}) \tag{8}$$

with $\alpha = 4\pi k/\lambda$ the absorption coefficient and $T(\lambda)$ the 3D transmission.

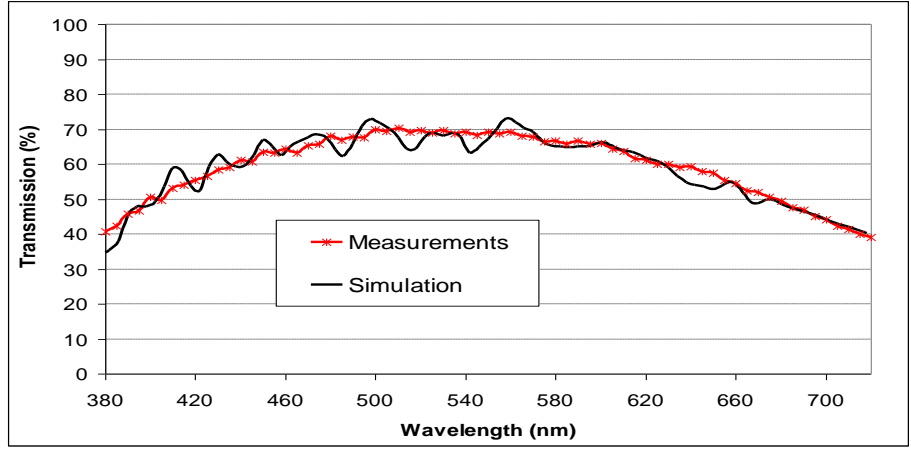

Fig. 12. Transmission simulated and Quantum Efficiency measured at normal incidence for the reduced stack

### 5.3 Sensor optimization

Firstly, sensors with a 1.75μm pixel pitch and a standard stack have been processed with and without the optimized antireflective coating. We have measured the Quantum Efficiency with an aperture of f#=8 to be close to a collimated beam (same conditions as the simulation results presented on Figure 9). Result in the green range (532±10nm) is shown on Table 2. We measure a gain of about 1.07 like we have obtained in simulation.

The second optimization consists in the reduction of the optical stack of about 30% by removing layers. We have measured the Quantum Efficiency with an aperture of f#=2.8 to be in the same condition that the final product. Moreover effect of stack reduction is more clearly seen with high incident angles than with a collimated beam. Result is shown on Table 2. We measure a gain of about 1.18 in the green range (532±10nm) for this aperture, very close to the gain that we observed in our simulation showing the relevance of our 3D narrowband simulation model. Finally, we observed in measure almost the same Quantum Efficiency response for the reduced stack for a large aperture range (f#=2, 2.8, and 8) meaning that light collection is very efficient with this new process.

Table. 2. Gain comparison in the green range between simulation and measurements for two sensor optimization process

| Green Gain | Simulation | Measurements |
|---|---|---|
| **Antireflective coating** | 1.08 | 1.07 |
| **Stack reduction** | 1.15 | 1.18 |

## 6. CONCLUSION

In this paper, we have presented the methodology used to simulate rigorously the optical performances of our CMOS image sensor, using FDTD software. We have shown the limitations of One-Dimensional and Two-Dimensional simulations for that kind of applications where pixel size shrink leads to asymmetric architecture, and the need of Three-Dimensional modeling. We have also shown that monochromatic simulations are not efficient due to oscillations observed in the optical stack transmission, and the need of broadband or narrowband simulations to obtain accurate results. Finally, we have proposed two applications for optical performance optimization of the sensor: an antireflective coating at photodiode interface, and a reduction of the stack height. We have compared the Quantum Efficiency measurements to the results of our simulations, showing the same optical gain and thus proving the relevance of our optical simulation modeling.

## ACKNOWLEDGMENT

The authors would like to thank James Pond from Lumerical Solutions for his support in this investigation.